\documentclass[aps,reprint,superscriptaddress,amsmath,amssymb,amsfonts,prb]{revtex4-1}
\usepackage{xcolor}
\definecolor{darkgreen}{rgb}{0,0.6,0}
\definecolor{cyan}{rgb}{0,0.7,0.8}
\usepackage[colorlinks=true,citecolor=darkgreen,urlcolor=cyan]{hyperref}

\usepackage{graphicx}
\usepackage[caption=false]{subfig}
\usepackage{braket}
\usepackage{epstopdf} 
\begin{document}

%\preprint{APS/123-QED}

\title{Linear classifier, least-squares cost function, and outliers}

% Force line breaks with \\

\author{Babatunde M. Ayeni}
\email{babatunde.ayeni@students.mq.edu.au}
\affiliation{Department of Physics \& Astronomy, Macquarie University, NSW 2109, Australia}
%\altaffiliation[Also at ]{Centre for Engineered Quantum Systems, Dept. of Physics \& Astronomy, Macquarie University, NSW 2109, Australia.}
%\affiliation{%


\date{August 28, 2018}

\pacs{Valid PACS appear here}% PACS, the Physics and Astronomy
%\keywords{Robust statistics, Least squares, outlier}
%Use showkeys class option if keyword

\begin{abstract}
This is a set of introductory notes on the subject of data classification using a linear classifier and least-squares cost (or “error”) function, and the negative effect of the presence of outliers on the decision boundary of the linear discriminant. We also show how a simple scaling could make the outlier less significant, thereby obtaining a much better decision boundary. We present some numerical results.
\end{abstract}

\maketitle
%\tableofcontents

\section{Introduction}
Computational statistics and machine learning are two different but closely related fields with, usually, {similar} methods. {For example, regression analysis and data classification are two problems that are common to both fields: on the one hand, regression analysis involves finding a mathematical model from a given set of training data such that one could make prediction of the output of a different input. On the other hand, data classification, from a supervised learning perspective, involves finding a model that \emph{learns} the classification of input training data with the aim of predicting the class (output) of any other \emph{similar} test data. Hence, the two problems quite similar, with the major difference being that \emph{labeling} is discrete in data classification 
but continuous in regression analysis. Therefore, it is possible to give the two problems a unified treatment.}

In order to obtain the model, we define a cost function that minimizes the distance between the data and the model. An often easy choice is sum-of-squares cost function, commonly referred to as \emph{least squares}. This cost function is most suitable to data with normal distribution, but also sometimes applied to data which are not normal (usually at the expense of accuracy). This choice is motivated because of the many analytical properties least squares method enjoys and the ease of its implementation. 

However, it is known that least squares is not robust against outliers. Several notable attempts have been made on finding solutions to this problem, see Ref. \onlinecite{WikiRobustStat} (and references therein) for historical facts. These led to the development of other methods that are referred to as ``least squares alternatives'' to resolve the \emph{outlier problem}, though they are either computationally inefficient or have some other limitations. Nonetheless, the \emph{outlier problem} with \emph{least squares} has not been resolved, and for the sake of distinction, it is now called \emph{ordinary least squares} (oLS) to differentiate it from the other ``least squares alternatives.'' 

There is no mathematically rigorous definition of what deserves to be called an outlier in a data. A common idea is that a data point is called an outlier if it does not follow the pattern of the remaining data points. Loosely speaking, if some set of data points are unusually ``far away'' from the expected region of majority of the data points, the ``strayed'' data points may be called outliers. Outliers can arise due to many reasons, including experimental errors either due to faulty equipments, imprecise set up, or environmental conditions; human error; or forging of results. Except where outliers are due to a known cause, they should be retained in a data, and rather use a statistical method that is robust against outliers. {When present in data, outliers may render the decision boundary sub-optimal, hence giving a poorer accuracy on classification.}

In this article, we review linear classifier using a least squares cost function which takes into account the possibility of data with outliers, and we discuss how the effect of the outliers may be reduced using a simple idea of \emph{scaling or re-weighting} the training input data. We apply the method to a synthetic dataset in two dimensions for binary classification, and also to one ``real world'' dataset---MNIST dataset for handwriting recognition---for multiple classification.

\section{Linear classifier using least squares error}\label{Sec:LinClassReview}
A linear classifier is a tool in machine learning that is often used for quick data exploration because it is fast to train and very easy to implement, albeit at the expense of its accuracy. It {can} become a competitive method if the dimensionality of input space is very high. An example problem where linear classifier could work quite well is document classification. For a more detailed review of linear classifiers, see Ref.~\onlinecite{Yuan2012}.

The aim of statistical classification, in general, is to assign a given data into one of many classes. Given a data represented as $(\mathbf{x}, t)$, where $\mathbf{x}$ is the vector representation of the data and $t$ is the corresponding label, the goal is to classify $\mathbf{x}$ into one of $K$ number of classes $\mathcal{C}^{(k)}$, where $k=1, \ldots, K$, using its label $t$ (in the case where each input data belongs to only one class). In cases where data is linearly separable, a linear classifier is sufficient, otherwise one should resort to one of the nonlinear methods such as neural network strategies.\citep{Bishop2006}

\subsection{Binary classification}
Our review shall mainly focus on binary classification, where there are two classes of data: $\mathcal{C}^{(1)}$ and $\mathcal{C}^{(2)}$ (i.e. $K=2$), without loss of generalization to multiple classification.

The usual starting point to binary classification is to construct a linear discriminant 
\begin{equation}
y(\mathbf{x}) = \mathbf{w}^{\mathrm{T}}\mathbf{x} + w_0, \label{Eq:DiscriminantFunction1}
\end{equation}
where $\mathbf{w}$ is the weight vector, $w_0$ is the bias (i.e. negative of the threshold), $\mathbf{x}$ is the input data vector. Hereafter,  $y(\mathbf{x})$ will be called \emph{ordinary linear discriminant} (oLD) in order to differentiate it from another variable, \emph{scaled linear discriminant} (sLD), that we introduce later. It is convenient to write Eq.~\eqref{Eq:DiscriminantFunction1} as
\begin{equation}
y(\mathbf{x'}) = \mathbf{w'}^{\mathrm{T}}\mathbf{x'}, \label{Eq:DiscriminantFunction2}
\end{equation}
where $\mathbf{w'} = (w_0, \mathbf{w}^T)^T$ and  $\mathbf{x'} = (x_0, \mathbf{x}^T)^T$, with $x_0=1$ (a ``dummy'' variable). The new $\mathbf{w'}$ and $\mathbf{x'}$ are commonly referred to as augmented weight vector and augmented input vector, respectively. The input vector $\mathbf{x}$ is assigned to class $\mathcal{C}^{(1)}$ if the discriminant $y(\mathbf{x}) > 0 $ or to class $\mathcal{C}^{(2)}$ if $y(\mathbf{x}) < 0 $, while it is exactly on the decision boundary if $y(\mathbf{x}) = 0$. 

The aim is to learn the $\mathbf{w}'$ that does this classification from the available labelled training data,  $\mathcal{S} = \{\mathbf{x}_n, t_n \}$ where $n \in [1,N]$, $N$ is the total number of training data, $\mathbf{x}_n$ is the input data vector, and $t_n$ is the  target binary variable whose value {we choose to be} $+1$ or $-1$ depending on whether $\mathbf{x}_n$ belongs to class $\mathcal{C}^{(1)}$ or $\mathcal{C}^{(2)}$. We employ \emph{least squares} as the cost function, defined as
\begin{equation}
C(\mathbf{w'}) = \frac{1}{2} \sum_{n=1}^N (y(\mathbf{x'}_n) - t_n )^2. \label{Eq:CostFuncBin}
\end{equation}
{To obtain the weight vector we seek to minimize the distance between the given target value $t_n$ and the model's prediction $y(\mathbf{x}'_n)$}. It is known that the minimization of this cost function is equivalent to the maximization of the loglikelihood of a Gaussian probability distribution with respect to the weight vector and bias.\citep{Bishop2006} The least squares approach therefore implies the data have an assumed Gaussian distribution, which may not be true, as the distribution is not known a priori. This is one of the often cited reasons to discourage the use of least squares on data that are not normally distributed. 

By minimizing the cost function of Eq.~\eqref{Eq:CostFuncBin} with respect to $\mathbf{w}'$, the (augmented) weight vector can be derived to be
\begin{equation}
\mathbf{w'} = \left( \sum_n \mathbf{x'}_n  \mathbf{x'}_n^{\mathrm{T}} \right)^{-1} \left(\sum_n t_n \mathbf{x'}_n \right). \label{Eq:WeightVectorEqn}
\end{equation}

\section{The effect of outliers on the decision boundary}\label{Sec:ExpoOutlier}
\begin{figure}
\vspace{20pt}
\includegraphics[width=\columnwidth]{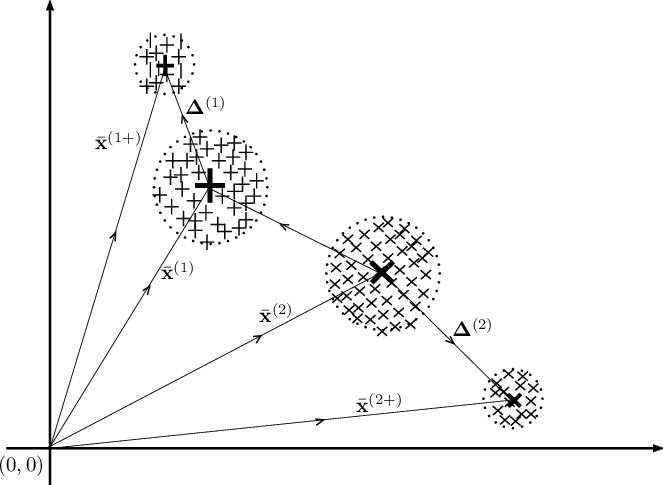}
\caption{Illustration of training data with two classes: ``pluses'' and ``crosses,'' represented here as \emph{ideal} circular clouds of data. Let the big clouds represent the  ``normal data'' and the small clouds some possible outliers. If there is no outlier, the number of points in the small clouds become zero. We let the collection of ``pluses'' be class $\mathcal{C}^{(1)}$ and the collection of ``crosses'' be class $\mathcal{C}^{(2)}$. The other variables are explained in the main text.} 
\label{Fig:OutlierGeneral}
\end{figure}

In this section, we show how the outlier problem manifest in least squares linear classifiers, where we assume that the input training data is ideal and normally distributed. For ease of visualization, we illustrate the problem in two dimensions as in Fig.~\ref{Fig:OutlierGeneral}, but this does not deduct from our  conclusion generally; The same conclusion holds true in higher dimensions (as we never use any feature peculiar to the dimensionality of the input space, except as an aid for visualization).

We assume that the data in each class is split into ``normal'' data and a possible outlier data. Let the number of ``normal'' data points be $N^{(k)}$ and the number of outliers be $N^{(k+)}$ in each class $\mathcal{C}^{(k)}$, where $k=1,2$. In the case where there is no outlier in a class, $N^{(k+)} = 0$ for that class.

We have shown in the previous section that the weight vector is determined from Eq.~\ref{Eq:WeightVectorEqn}, where the input data vector $\mathbf{x'}_n$ is the augmented vector, but in the following we will drop the primes for convenience. 

Because the class that each input data belongs to during training is known, then rather than indexing the input vector with a single index $n$, we use a two-index notation $(k, u_k)$, where $k$ indexes the class $\mathcal{C}^{(k)}$ and $u_k$ indexes the sample that belongs to that class. (We index the outlier with $m_k$.) We split the input data in each class into a reference vector---taken to be the mean vector $\bar{\mathbf{x}}$---and a ``residual vector'' $\mathbf{\varepsilon}$. As in Fig.~\ref{Fig:OutlierGeneral}, let $\bar{\mathbf{x}}^{(1)}$ and $\bar{\mathbf{x}}^{(2)}$ be the mean vectors of the ``big clouds,''  and $\bar{\mathbf{x}}^{(1+)}$ and $\bar{\mathbf{x}}^{(2+)}$ be the mean vectors of the ``small clouds'' (which are assumed here to be the outliers), and $\mathbf{\Delta}^{(1)}$, $\mathbf{\Delta}^{(2)}$ be the displacement vectors of the ``small clouds'' from the ``big clouds.'' Therefore, the ``normal'' input data and outlier can be written respectively as
\begin{align}
&\mathbf{x}^{(k)}_{u_k} = \bar{\mathbf{x}}^{(k)} + \mathbf{\varepsilon}^{(k)}_{u_k}, \label{Eq:DataRep1}\\
&\mathbf{x}^{(k+)}_{m_k} = \bar{\mathbf{x}}^{(k+)} + \mathbf{\varepsilon}^{(k+)}_{m_k}. \label{Eq:DataRep2}
\end{align}
For convenience, we split the weight vector Eq.~\ref{Eq:WeightVectorEqn} into parts as $\mathbf{w} = I S$, where 
\begin{equation}
I =  \left( \sum_n \mathbf{x}_n  \mathbf{x}_n^{\mathrm{T}} \right)^{-1}
\end{equation}
and 
\begin{equation}
S =  \left(\sum_n t_n \mathbf{x}_n \right).
\end{equation}

We now write $S$ and $I$ in terms of Eqs.~\ref{Eq:DataRep1} and \ref{Eq:DataRep2}. Starting with $S$:
\begin{align}
S = \sum_k \left(\sum_{u_k} t^{(k)}_{u_k} \mathbf{x}^{(k)}_{u_k}  + \sum_{m_k} t^{(k+)}_{m_k} \mathbf{x}^{(k+)}_{m_k}  \right),
\end{align}
which is split into sum over the ``normal'' data and the possible 	outlying data. 
But as the target variable $t$ only depends on the class, $t^{(k)}_{u_k} = t^{(k)}$, for all $u_k$ and $m_k$, 
\begin{align}
S = \sum_k t^{(k)} \left(\sum_{u_k}  \mathbf{x}^{(k)}_{u_k}  + \sum_{m_k} \mathbf{x}^{(k+)}_{m_k}  \right).
\end{align}
Substitute Eqs.~\ref{Eq:DataRep1} and \ref{Eq:DataRep2} into the above equation to give
\begin{align}
S = \sum_k t^{(k)} \left[\sum_{u_k}  \left( \bar{\mathbf{x}}^{(k)} + \mathbf{\varepsilon}^{(k)}_{u_k} \right) + \sum_{m_k} \left(  \bar{\mathbf{x}}^{(k+)} + \mathbf{\varepsilon}^{(k+)}_{m_k}  \right) \right].
\end{align}
For the assumed symmetric cloud (i.e. the ideal case), the sum over the relative vectors, $\sum_{u_k} \mathbf{\varepsilon}^{(k)}_{u_k} = 0$, in all cases. Therefore,
\begin{equation}
S =  \sum_k t^{(k)} \left[ N^{(k)}\bar{\mathbf{x}}^{(k)}  + N^{(k+)}\bar{\mathbf{x}}^{(k+)} \right].
\end{equation}

We now consider the $I$ term.
\begin{align}
I = \left[\sum_k \left(\sum_{u_k} \mathbf{x}^{(k)}_{u_k}\mathbf{x}^{(k)^{\mathrm{T}}}_{u_k}  + \sum_{m_k} \mathbf{x}^{(k+)}_{m_k} \mathbf{x}^{(k+)^{\mathrm{T}}}_{m_k}   \right)\right]^{-1}.
\end{align}
Again, we substitute Eqs.~\ref{Eq:DataRep1} and \ref{Eq:DataRep2} into the above equation to get
\begin{align}
I & = \left\{ \sum_k \left[ \sum_{u_k} \left( \bar{\mathbf{x}}^{(k)} \bar{\mathbf{x}}^{(k)^{\mathrm{T}}}   +  \mathbf{\varepsilon}^{(k)}_{u_k} \mathbf{\varepsilon}^{(k)^{\mathrm{T}}}_{u_k} \right)  \right. \right.\nonumber \\
& \qquad \qquad \qquad  +  \left. \left.  \sum_{m_k} \left( \bar{\mathbf{x}}^{(k+)} \bar{\mathbf{x}}^{(k+)^{\mathrm{T}}} + \mathbf{\varepsilon}^{(k+)}_{m_k} \mathbf{\varepsilon}^{(k+)^{\mathrm{T}}}_{m_k}  \right) \right] \right \}^{-1},
\end{align}
which simplifies to 
\begin{align}
I = \left[\sum_k \left( N^{(k)} \bar{\mathbf{x}}^{(k)} \bar{\mathbf{x}}^{(k)^{\mathrm{T}}} + N^{(k+)} \bar{\mathbf{x}}^{(k+)} \bar{\mathbf{x}}^{(k+)^{\mathrm{T}}} \right) \right. \nonumber \\
\qquad \qquad \left. + \sum_k \left( \sum_{u_k} \mathbf{\varepsilon}^{(k)}_{u_k} \mathbf{\varepsilon}^{(k)^{\mathrm{T}}}_{u_k} +  \sum_{m_k} \mathbf{\varepsilon}^{(k+)}_{m_k} \mathbf{\varepsilon}^{(k+)^{\mathrm{T}}}_{m_k}  \right) \right]^{-1}.
\end{align}
We shall let 
\begin{equation}
M = \sum_k \left( N^{(k)} \bar{\mathbf{x}}^{(k)} \bar{\mathbf{x}}^{(k)^{\mathrm{T}}} + N^{(k+)} \bar{\mathbf{x}}^{(k+)} \bar{\mathbf{x}}^{(k+)^{\mathrm{T}}} \right),
\end{equation}
and 
\begin{equation}
E = \sum_k \left( \sum_{u_k} \mathbf{\varepsilon}^{(k)}_{u_k} \mathbf{\varepsilon}^{(k)^{\mathrm{T}}}_{u_k} +  \sum_{m_k} \mathbf{\varepsilon}^{(k+)}_{m_k} \mathbf{\varepsilon}^{(k+)^{\mathrm{T}}}_{m_k}  \right).
\end{equation}
Therefore,
\begin{equation}
I = (M + E)^{-1}.
\end{equation}
Since we do not know the exact nature of $E$, the series expansion of the above expression cannot be guaranteed, but for the purpose of ``exposition,'' we make the crude approximation for $I$ to the leading term, hence
\begin{equation}
I \approx M^{-1}, 
\end{equation}
which involves only the mean vectors of the two classes. Written explicilty,
\begin{equation}
I \approx \left[ \sum_k \left( N^{(k)} \bar{\mathbf{x}}^{(k)} \bar{\mathbf{x}}^{(k)^{\mathrm{T}}} + N^{(k+)} \bar{\mathbf{x}}^{(k+)} \bar{\mathbf{x}}^{(k+)^{\mathrm{T}}} \right)\right]^{-1}.
\end{equation}

If we choose for the class labels the values $t^{(1)}=1$ and $t^{(2)}=-1$, the expressions for $S$ and $I$ can be  simplified to be 
\begin{align}
&S = N^{(1)} \bar{\mathbf{x}}^{(1)} +  N^{(1+)} \bar{\mathbf{x}}^{(1+)} -  N^{(2)} \bar{\mathbf{x}}^{(2)} - N^{(2+)} \bar{\mathbf{x}}^{(2+)}, \\
&I \approx \left( N^{(1)} \bar{\mathbf{x}}^{(1)} \bar{\mathbf{x}}^{(1)^{\mathrm{T}}} + N^{(1+)} \bar{\mathbf{x}}^{(1+)} \bar{\mathbf{x}}^{(1+)^{\mathrm{T}}}  \right. \nonumber \\
& \qquad \qquad \left. + \quad N^{(2)} \bar{\mathbf{x}}^{(2)} \bar{\mathbf{x}}^{(2)^{\mathrm{T}}} 
+ N^{(2+)} \bar{\mathbf{x}}^{(2+)} \bar{\mathbf{x}}^{(2+)^{\mathrm{T}}}  \right)^{-1}.
\end{align}
We can rewrite these expressions in term of densities $\rho^{(k)} = N^{(k)}/N$, where $N = \sum_k N^{(k)}$, as
\begin{align}
&S = N \left(\rho^{(1)} \bar{\mathbf{x}}^{(1)} + \rho^{(1+)} \bar{\mathbf{x}}^{(1+)} - \rho^{(2)}\bar{\mathbf{x}}^{(2)} - \rho^{(2+)} \bar{\mathbf{x}}^{(2+)} \right), \label{Eq:theS} \\
&I \approx N^{-1} \left( \rho^{(1)} \bar{\mathbf{x}}^{(1)} \bar{\mathbf{x}}^{(1)^{\mathrm{T}}} +  \rho^{(1+)}\bar{\mathbf{x}}^{(1+)} \bar{\mathbf{x}}^{(1+)^{\mathrm{T}}}  \right. \nonumber \\
& \qquad \qquad \left. + \quad \rho^{(2)} \bar{\mathbf{x}}^{(2)} \bar{\mathbf{x}}^{(2)^{\mathrm{T}}} 
+ \rho^{(2+)} \bar{\mathbf{x}}^{(2+)} \bar{\mathbf{x}}^{(2+)^{\mathrm{T}}}  \right)^{-1},
\end{align}
where now $\rho^{(k)}$ is the density of the ``normal data'' and $\rho^{(k+)}$ is the density of outliers in each class. From these equations we can explore different cases, namely, when there are no outliers and otherwise. Since the case without an outlier poses no problem, we will consider only the special case where there are some outlier data, speficically in the second class. (Similar conclusion can be reached for the other cases.)

\subsubsection{Outlier in class $\mathbf{C^{(2)}}$}
We now consider the example of when there is an outlier, e.g. in class $\mathcal{C}^{(2)}$. [The same conclusion holds true if we instead choose the outlier to be in class  $\mathcal{C}^{(1)}$]. To that end, we let $\rho^{(1+)} = 0$, and also consider the special values: $\rho^{(1)} = 1/2$, $\rho^{(2)} = \gamma$, $\rho^{(2+)} = \frac{1}{2} - \gamma$, where $0 < \gamma \le 1/2$.

Under these assumptions, $S$ and $I$ become
\begin{align}
& S = N \left[ \frac{1}{2} \bar{\mathbf{x}}^{(1)} -\gamma \bar{\mathbf{x}}^{(2)}  - \left(\frac{1}{2} - \gamma\right) \bar{\mathbf{x}}^{(2+)} \right], \label{Eq:SForCase3}\\
& I \approx N^{-1} \left[ \frac{1}{2} \bar{\mathbf{x}}^{(1)}\bar{\mathbf{x}}^{(1)^{\mathrm{T}}} + \gamma\bar{\mathbf{x}}^{(2)}\bar{\mathbf{x}}^{(2)^{\mathrm{T}}} \right. \nonumber \\
& \qquad \qquad \qquad + \left. \left(\frac{1}{2} - \gamma \right) \bar{\mathbf{x}}^{(2+)}\bar{\mathbf{x}}^{(2+)^{\mathrm{T}}} \right]^{-1}. \label{Eq:IForCase3}
\end{align}
Even though the density of both classes are balanced, the density $(1/2 - \gamma)$ of the outlier (at some value of $\gamma$, say near $2$) and its coordinate values, $\bar{\mathbf{x}}^{(2+)}$, can still exert a ``pulling force'' on the boundary, ``pulling'' it more towards the second class. In other words, the denser or farther the outlier is from the ``normal'' data, determined either through $\gamma$ or the position vector $\bar{\mathbf{x}}^{(2+)}$, the more the boundary line is ``pulled'' towards the half plane of the outliers, and thereby gives a poorer result on classification. This effect can be easily seen in any of the above equations. Therefore, increasing either the density or the position of the outliers can accentuate their effect. (We assume, as in a real scenario, that the density of the outliers is less than the density of the ``normal'' data---otherwise the outliers should rather be considered as the ``normal'' data).

\section{Making \emph{outliers} less significant} \label{Sec:SolutionToOutlier}
Having reviewed how outliers pose to be a problem in \emph{least squares} error function, {we now review a simple means through which they can be made less significant: \emph{by applying some length scale to the discriminant function}.} 

We start with the realization that, for example, in binary classification, the \emph{classification criterion} is the sign of the discriminant function
\begin{equation}
y(\mathbf{x}) = \mathbf{w}^{\mathrm{T}}\mathbf{x} + w_0, \label{Eq:DiscriFuncAgain}
\end{equation}
where $y(\mathbf{x}) < 0$ if $\mathbf{x}$ is in the lower plane, $y(\mathbf{x}) = 0$ if $\mathbf{x}$ is on the decision boundary, and $y(\mathbf{x}) > 0$ if $\mathbf{x}$ is in the upper plane. In multiple classification using linear discriminant, the \emph{classification criterion} is the maximum value of the discriminant function of the different classes. Common with both classification problem is that the \emph{classification criterion} is ``scale invariant.'' Basically, this means that if we divide $y(\mathbf{x})$ by some length scale $S$, the 
\emph{classification criterion} of the discriminant does not change, though its value changes; In binary classification, the sign of $y(\mathbf{x})$ is unchanged by a length scale. Also, in multiple classification, all the entries of $y(\mathbf{x})$ are scaled uniformly, so that the highest number remains so. 

For reasons that will be clear later, we work with the augmented version of the above equation,
\begin{equation}
y(\mathbf{x'}) = \mathbf{w'}^{\mathrm{T}}\mathbf{x'}, \label{Eq:DiscriFuncAgain}
\end{equation}
where $\mathbf{x'}$ is the augmented input vector, with length $\Vert \mathbf{x'} \Vert = \sqrt{ 1 + \Vert \mathbf{x} \Vert^2 }$. We define the scaled version of the equation above as
\begin{equation}
Y(\mathbf{x'}) := \frac{y(\mathbf{x'})}{\Vert \mathbf{x'} \Vert} =  \frac{\mathbf{w'}^{\mathrm{T}}\mathbf{x'} }{\Vert \mathbf{x'} \Vert} = \mathbf{w'}^{\mathrm{T}} \mathbf{X'},
\end{equation}
where $\mathbf{X'} = \frac{\mathbf{x'}}{\Vert \mathbf{x'} \Vert}$ is the scaled augmented input vector. Using this scaled version, all the formulas of the \emph{least squares} linear classifiers, namely, the cost function and the expression for the weight vector, remain the same, but the input vectors are now scaled.

We posit that using this scaled version will ``alleviate'' the \emph{outlier problem}. To show this, we recall the expressions Eqs.~\ref{Eq:SForCase3} and \ref{Eq:IForCase3} of the example with outlier in Sec.~\ref{Sec:ExpoOutlier}. The scaled versions are given as 
\begin{align}
& S = N \left[ \frac{1}{2} \bar{\mathbf{X}}^{(1)} -\gamma \bar{\mathbf{X}}^{(2)}  - \left(\frac{1}{2} - \gamma\right) \bar{\mathbf{X}}^{(2+)} \right], \label{Eq:SForOutlierSol}\\
& I \approx N^{-1} \left[ \frac{1}{2} \bar{\mathbf{X}}^{(1)}\bar{\mathbf{X}}^{(1)^{\mathrm{T}}} + \gamma\bar{\mathbf{X}}^{(2)}\bar{\mathbf{X}}^{(2)^{\mathrm{T}}} \right. \nonumber \\
& \qquad \qquad \qquad + \left. \left(\frac{1}{2} - \gamma \right) \bar{\mathbf{X}}^{(2+)}\bar{\mathbf{X}}^{(2+)^{\mathrm{T}}} \right]^{-1}, \label{Eq:IForOutlierSol}
\end{align}
where for typographical convenience we have also dropped the primes on the augmented vectors.

We simplify $S$ and $I$. Starting with $S$,
\begin{equation}
S = N \left[ \frac{1}{2} \frac{\bar{\mathbf{x}}^{(1)}}{\Vert \bar{\mathbf{x}}^{(1)} \Vert} -\gamma \frac{\bar{\mathbf{x}}^{(2)}}{\Vert \bar{\mathbf{x}}^{(2)} \Vert}  - \left(\frac{1}{2} - \gamma\right) \frac{\bar{\mathbf{x}}^{(2+)}}{\Vert \bar{\mathbf{x}}^{(2+)} \Vert} \right]. 
\end{equation}
We assume the length of the outlier, $\Vert \bar{\mathbf{x}}^{(2+)} \Vert$, is higher than non-outlier vectors, and hence factor it out. Therefore,
\begin{align}
S & = \frac{N}{\Vert \bar{\mathbf{x}}^{(2+)} \Vert} \left[ \frac{1}{2} \Vert \bar{\mathbf{x}}^{(2+)} \Vert \frac{\bar{\mathbf{x}}^{(1)}}{\Vert \bar{\mathbf{x}}^{(1)} \Vert} - \gamma \Vert \bar{\mathbf{x}}^{(2+)} \Vert \frac{\bar{\mathbf{x}}^{(2)}}{\Vert \bar{\mathbf{x}}^{(2)} \Vert}  \right. \nonumber \\
& \qquad \qquad \qquad \qquad \qquad - ~~ \left. \left(\frac{1}{2} - \gamma\right) \bar{\mathbf{x}}^{(2+)} \right].
\end{align}

%Let $S = \frac{N}{\Vert \bar{\mathbf{x}}^{(2+)} \Vert} \tilde{S}$, where

Similarly, we simplify $I$,
\begin{align}
& I \approx N^{-1} \left[ \frac{1}{2} \frac{\bar{\mathbf{x}}^{(1)} \bar{\mathbf{x}}^{(1)^{\mathrm{T}}}}{\Vert \bar{\mathbf{x}}^{(1)} \Vert^2} + \gamma \frac{\bar{\mathbf{x}}^{(2)}\bar{\mathbf{x}}^{(2)^{\mathrm{T}}} }{\Vert \bar{\mathbf{x}}^{(2)} \Vert^2} \right. \nonumber \\
& \qquad \qquad \qquad + \left. \left(\frac{1}{2} - \gamma \right) \frac{\bar{\mathbf{x}}^{(2+)}\bar{\mathbf{x}}^{(2+)^{\mathrm{T}}}}{\Vert \bar{\mathbf{x}}^{(2+)} \Vert^2} \right]^{-1},
\end{align}
which we write as
\begin{align}
& I \approx \frac{\Vert \bar{\mathbf{x}}^{(2+)} \Vert}{N} \left[ \frac{1}{2} \Vert \bar{\mathbf{x}}^{(2+)} \Vert \frac{\bar{\mathbf{x}}^{(1)} \bar{\mathbf{x}}^{(1)^{\mathrm{T}}}}{\Vert \bar{\mathbf{x}}^{(1)} \Vert^2}   + \gamma \Vert \bar{\mathbf{x}}^{(2+)} \Vert \frac{\bar{\mathbf{x}}^{(2)}\bar{\mathbf{x}}^{(2)^{\mathrm{T}}}}{\Vert \bar{\mathbf{x}}^{(2)} \Vert^2} \right. \nonumber \\
& \qquad \qquad \qquad \qquad + \left. \left(\frac{1}{2} - \gamma \right) \frac{\bar{\mathbf{x}}^{(2+)}\bar{\mathbf{x}}^{(2+)^{\mathrm{T}}}}{\Vert \bar{\mathbf{x}}^{(2+)} \Vert} \right]^{-1}.
\end{align}

%Let $I = \Vert \bar{\mathbf{x}}^{(2+)} \Vert N^{-1} \tilde{I}$, where

The weight vector $\mathbf{w}$ is 
\begin{align}
&\mathbf{w} \approx IS = \tilde{I} \tilde{S}, 
\end{align}
where
\begin{align}
\tilde{S} & = \left[ \frac{1}{2} \Vert \bar{\mathbf{x}}^{(2+)} \Vert \frac{\bar{\mathbf{x}}^{(1)}}{\Vert \bar{\mathbf{x}}^{(1)} \Vert} - \gamma \Vert \bar{\mathbf{x}}^{(2+)} \Vert \frac{\bar{\mathbf{x}}^{(2)}}{\Vert \bar{\mathbf{x}}^{(2)} \Vert}  \right. \nonumber \\
& \qquad \qquad \qquad \qquad \qquad - ~~ \left. \left(\frac{1}{2} - \gamma\right) \bar{\mathbf{x}}^{(2+)} \right],
\end{align}
and 
\begin{align}
&\tilde{I} = \left[ \frac{1}{2} \Vert \bar{\mathbf{x}}^{(2+)} \Vert \frac{\bar{\mathbf{x}}^{(1)} \bar{\mathbf{x}}^{(1)^{\mathrm{T}}}}{\Vert \bar{\mathbf{x}}^{(1)} \Vert^2}   + \gamma \Vert \bar{\mathbf{x}}^{(2+)} \Vert \frac{\bar{\mathbf{x}}^{(2)}\bar{\mathbf{x}}^{(2)^{\mathrm{T}}}}{\Vert \bar{\mathbf{x}}^{(2)} \Vert^2} \right. \nonumber \\
& \qquad \qquad \qquad \qquad + \left. \left(\frac{1}{2} - \gamma \right) \frac{\bar{\mathbf{x}}^{(2+)}\bar{\mathbf{x}}^{(2+)^{\mathrm{T}}}}{\Vert \bar{\mathbf{x}}^{(2+)} \Vert} \right]^{-1},
\end{align}
without the prefactors in front of $S$ and $I$. The most important thing from these equations is that the vectors of the ``normal'' data in both $\tilde{S}$ and $\tilde{I}$ have been ``scaled up'' by the length of the outlier thereby making them as equally important as, or even more important than, the outlier in determining the decision boundary. As a matter of fact, the farther the outlier the lesser its effect on the decision boundary. It is noteworthy that the outliers are still present in $\tilde{S}$ and $\tilde{I}$ (which is good, since they are originally part of the input data), but the decision boundary is now less sensitive to them. That is, we made the outliers less significant.  

We have shown how to improve statistical classification using \emph{least squares} in the presence of outliers, by scaling the augmented input data vector by its length. Though this shown through an example, but the idea applies generally. Not only is this also true approximately, but also in the exact case.

We present numerical proofs in Sec.~\ref{Sec:NumericalResults} that the method does what is expected. In our numerical implementation, we do not use the approximate equations but the exact equation for determining the weight vector as in Eq.~\ref{Eq:WeightVectorEqn}, only now scaling the input data vector as have been shown.

\subsection{Potential pitfall of the proposed solution and a possible solution}
In an attempt to  alleviate the \emph{outlier problem}, we proposed that the norm of the augmented input vector $\mathbf{x}'$ should be used as the scale without any justification. We give one now. It is tempting to want to use the norm  $\Vert \mathbf{x} \Vert $ of the original input vector $\mathbf{x}$ as the scale. However, this has a serious potential problem. For definiteness and simplicity, we consider the input data to be in two dimensions. Assuming that we scale with $\Vert \mathbf{x} \Vert$, then the scaled augmented input vector becomes
\begin{equation}
\mathbf{X}' = \frac{\mathbf{x}'}{\Vert \mathbf{x} \Vert},
\end{equation}
where $\mathbf{x}' = (1, x, y)^{\mathrm{T}}$, where $1$ is the ``dummy'' number, $x$ and $y$ are the coordinates of the input data vector $\mathbf{x}$. Therefore, the coordinates have the scaling transformation
\begin{align}
x \rightarrow \frac{x}{\sqrt{x^2 + y^2}}, \\
y \rightarrow \frac{y}{\sqrt{x^2 + y^2}},
\end{align}
which maps a point $(x,y)$ on the Cartesian plane to a point on a unit circle, as $x^2 + y^2=1$. This is dangerous if two different data points in two different classes are related by a positive scale factor. Under the above mapping, they map to the same point on the unit circle, and hence become non-separable. 
Specifically, let $\mathbf{x}_1$ and $\mathbf{x}_2$ be two distinct input that belong to two different classes in the original input space and related as $\mathbf{x}_2 = \lambda \mathbf{x}_1$, where $\lambda$ is some positive scale factor. We have that (for the scaled version of the original input vector)
\begin{equation}
\mathbf{X}_2 = \frac{\mathbf{x}_2}{\Vert \mathbf{x}_2 \Vert} = \frac{\lambda \mathbf{x}_1}{\Vert \lambda \mathbf{x}_1 \Vert} = \frac{\mathbf{x}_1}{\Vert  \mathbf{x}_1 \Vert} = \mathbf{X}_1.
\end{equation}
Therefore, different points on the plane now map to the same point on the unit circle. The scaled versions of the augmented input vectors are
\begin{align}
\mathbf{X}_1' = 
\begin{pmatrix}
\frac{1}{\sqrt{x_1^2+y_1^2}} \\ \frac{x_1}{\sqrt{x_1^2+y_1^2}} \\ \frac{y_1}{\sqrt{x_1^2 + y_1^2}}
\end{pmatrix}
\qquad 
\mathbf{X}_2' = 
\begin{pmatrix}
\frac{1}{\lambda} \frac{1}{\sqrt{x_1^2+y_1^2}} \\ \frac{x_1}{\sqrt{x_1^2+y_1^2}} \\ \frac{y_1}{\sqrt{x_1^2 + y_1^2}}
\end{pmatrix}.
\end{align}
With this, the usual assumed \emph{contiguity hypothesis} in linearly separable data may not be maintained as a data point may ``jump'' from a region of one class to a \emph{region} of another class under this transformation and might become not linearly separable anymore.

A solution that potentially resolves this problem is to use the norm of the augmented input vector $\mathbf{x}'$ as the length scale, with the length as $\Vert \mathbf{x'} \Vert = \sqrt{1 + x^2 + y^2}$. It is obvious that if $\mathbf{x}_2 = \lambda \mathbf{x}_1$, $\mathbf{x}_2' = (1, \lambda \mathbf{x}_1^{\mathrm{T}})^{\mathrm{T}}$, and $\Vert \mathbf{x}_2' \Vert \ne |\lambda| \Vert \mathbf{x}_1' \Vert$, and hence the scaled input vector $\mathbf{X}_1$ and $\mathbf{X}_2$, using the norm of the augmented input vector as the length scale, will not be equal. Furthermore, the scaled augmented vectors are
\begin{align}
\mathbf{X}_1' = 
\begin{pmatrix}
\frac{1}{\sqrt{1+ x_1^2+y_1^2}} \\ \frac{x_1}{\sqrt{1+ x_1^2+y_1^2}} \\ \frac{y_1}{\sqrt{1+ x_1^2 + y_1^2}}
\end{pmatrix},
\qquad 
\mathbf{X}_2' = 
\begin{pmatrix}
\frac{1}{\sqrt{1+ \lambda^2 x_1^2 + \lambda^2 y_1^2}} \\ \frac{\lambda x_1}{\sqrt{1 + \lambda^2 x_1^2 + \lambda^2 y_1^2}} \\ \frac{\lambda y_1}{\sqrt{1 + \lambda^2 x_1^2 + \lambda^2 y_1^2}}
\end{pmatrix},
\end{align}
which cannot be made equal by any value of $\lambda$ (except the trivial value of $\lambda=1$). Therefore, the \emph{contiguity hypothesis} is maintained.

However, the above solution also breaks down in the limit $\vert x \vert$, $\vert y \vert \gg 1$. There are two possible  solutions to this:
\begin{enumerate}
\item One can apply some uniform transformation to the input space, e.g. scaling or translation, to change the coordinate values to a range where addition of $1$ is significant. 

\item An alternative solution would be to use a length scale such as $\sqrt{c + x^2 + y^2}$, where $c >0$ is arbitrary, and such a value such that $\sqrt{c + x^2 + y^2} \not\approx \sqrt{x^2 + y^2}$.
\end{enumerate}
In our numerical tests, we have adopted the second suggestion with the value of $c=1$.

The above treatment generalizes to higher dimensional input space. The scaling transformation will map the coordinates of D-dimensional data points in the (assumed) Cartesian coordinate system to the surface of a hypersphere in D dimensions.

\section{Numerical Results}\label{Sec:NumericalResults}
We now provide numerical proofs that our proposed solution works efficiently. We test it both on synthetic data and ``real-world'' data, comparing the solutions to those obtained using ordinary linear discriminant. The real-world data is the MNIST dataset\cite{MNISTDataset} for handwriting image recognition. The synthetic data are randomly generated in two dimensions and made to look similar to Figure 4.4, Pg.~186 of Ref.~\onlinecite{Bishop2006}. The data contain two classes: ``red crosses'' and ``blue circles,'' with and without outlier. 

\begin{figure}
\includegraphics[clip,width=\columnwidth]{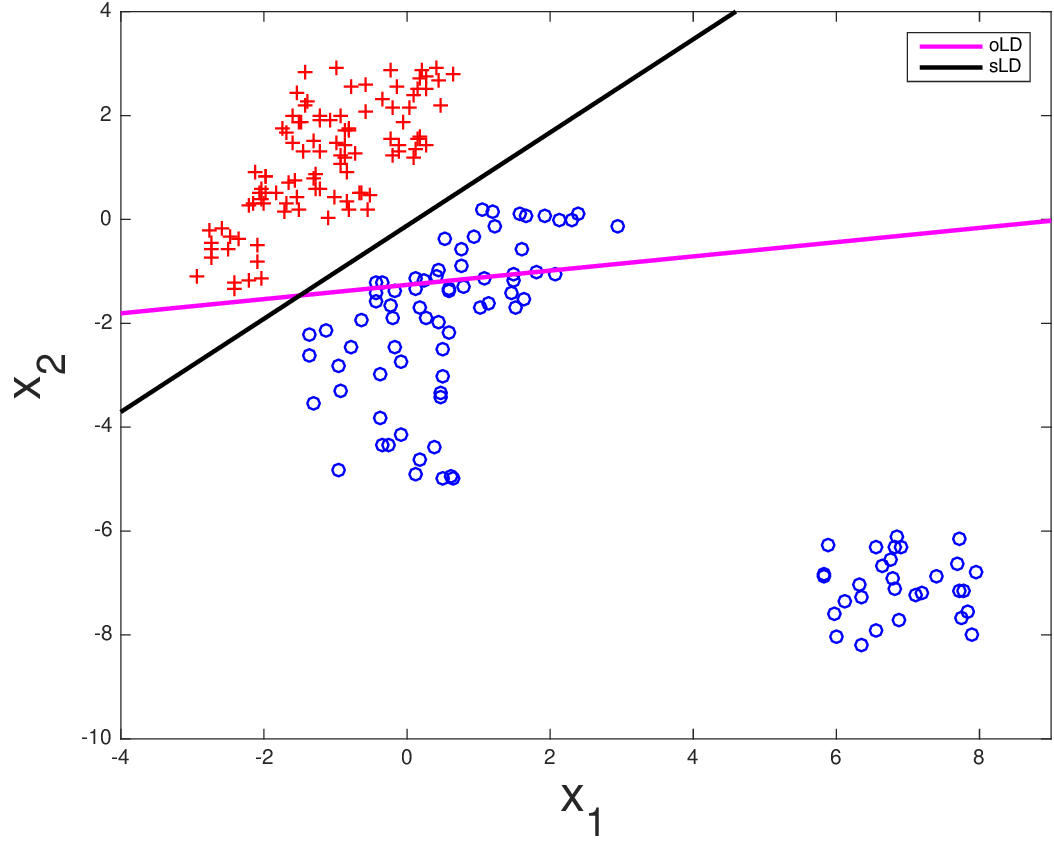}\\[0.3pt]
\caption{Random synthetic data in two dimensions with outliers. There are two classes: ``red crosses'' and ``blue circles.'' The magenta line is the decision boundary obtained using \emph{ordinary linear discriminant} (oLD), i.e. the conventional method, while the black line is the decision boundary obtained using our proposed method of \emph{scaled linear discriminant} (sLD).} 
\label{Fig:Outlier}
\end{figure}

First, we present results when the density population of data of the two classes are equal. In Fig.~\ref{Fig:Outlier}, there are 100 data points in both classes, i.e. 100 ``red crosses,'' 70 ``blue circles,'' and 30 ``blue circles'' outlier. The magenta line is the solution using \emph{ordinary} linear discriminant (oLD), i.e. using the conventional method, which, as is already known and also shown here again, fails in the presence of the outlier. The black line is our solution using \emph{scaled} linear discriminant (sLD), which gets the right boundary. 
While it is already known that oLD is sensitive to outlier, that is, not robust against outlier, our method of sLD is less sensitive to outlier, that is, robust against outlier, and moreover uses the same weight expression as the oLD but with scaled augmented input data vectors rather than ordinary augmented input data vectors. 
\begin{figure}
\includegraphics[clip,width=\columnwidth]{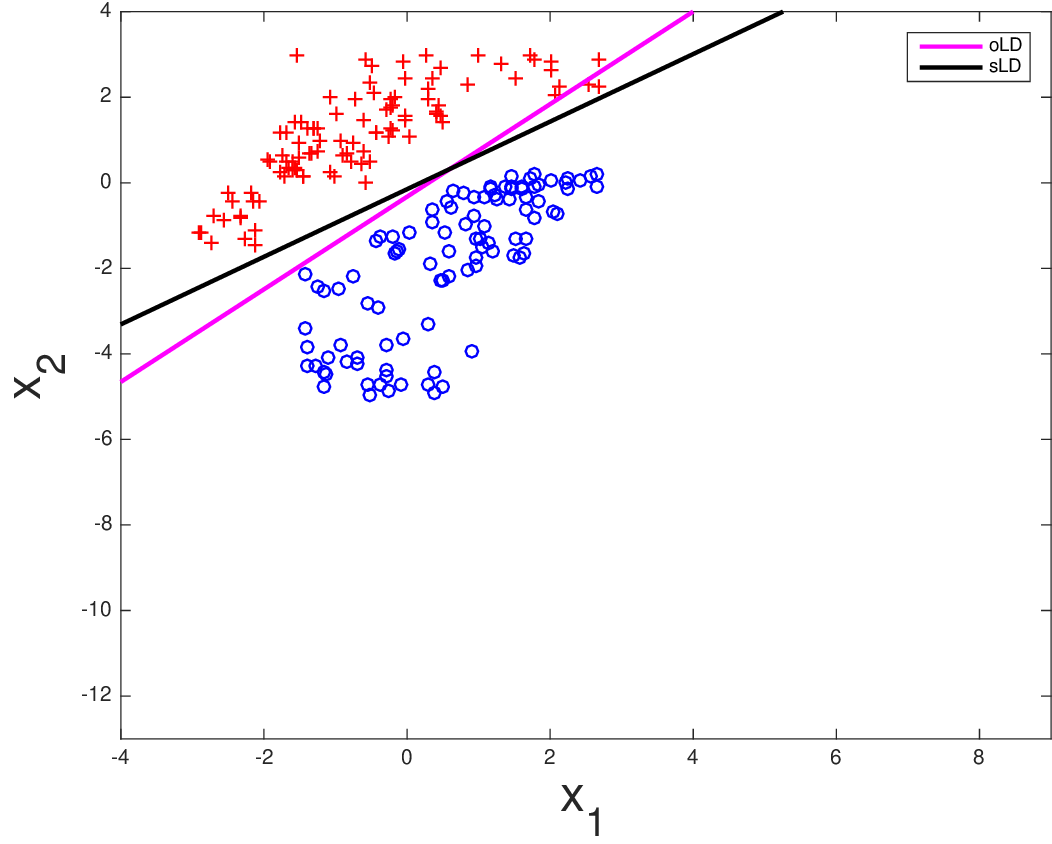} \\
\caption{Random synthetic data in two dimensions without outlier. The magenta line is the decision boundary obtained using oLD, while the black line is the decision boundary obtained using sLD. It can be seen that the result obtained using sLD is perhaps better than the result of oLD method.}
\label{Fig:Normal}
\end{figure}
We then also test the method in the ``normal'' case when there are no outliers, with 100 data points in each class. The method is as good, or perhaps better, than the conventional method as shown in Fig.~\ref{Fig:Normal}.

\begin{figure}
\includegraphics[clip,width=\columnwidth]{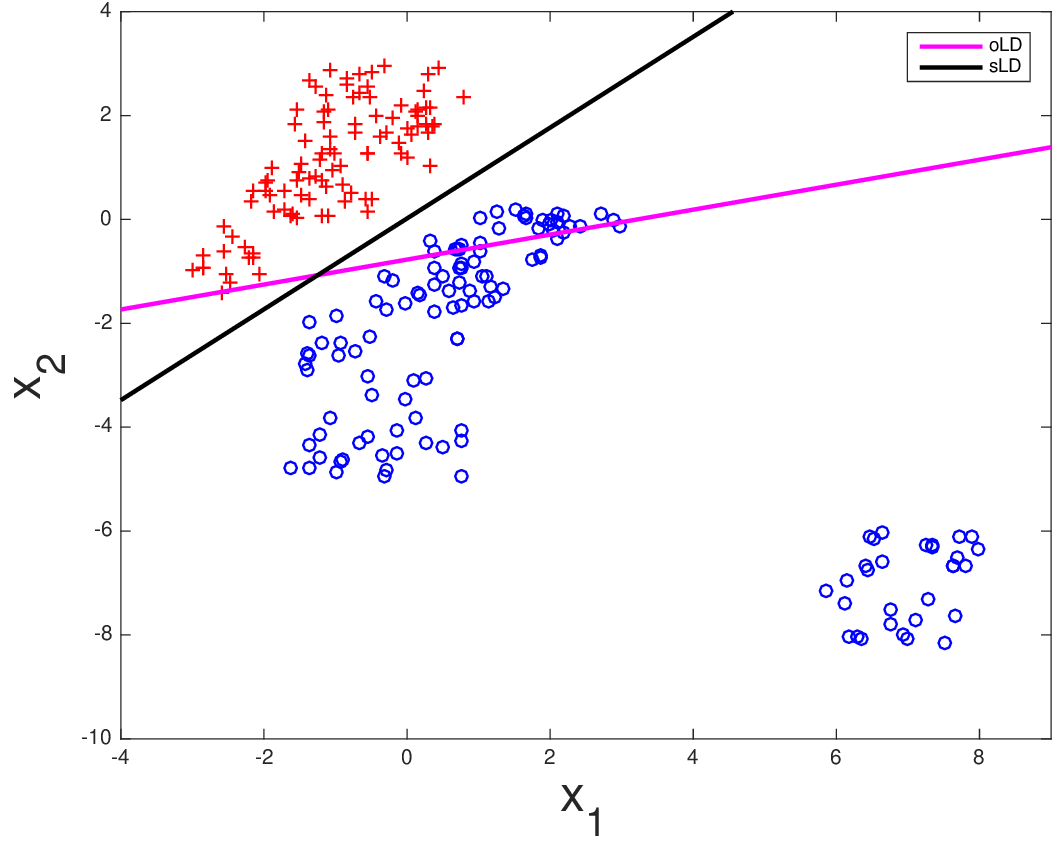}
\caption{Unequal density data population with outlier. Read text for more details.} 
\label{Fig:MoreDensityAndSamePosition}
\end{figure}

\begin{figure}
{\vspace{15pt}}
\includegraphics[clip,width=\columnwidth]{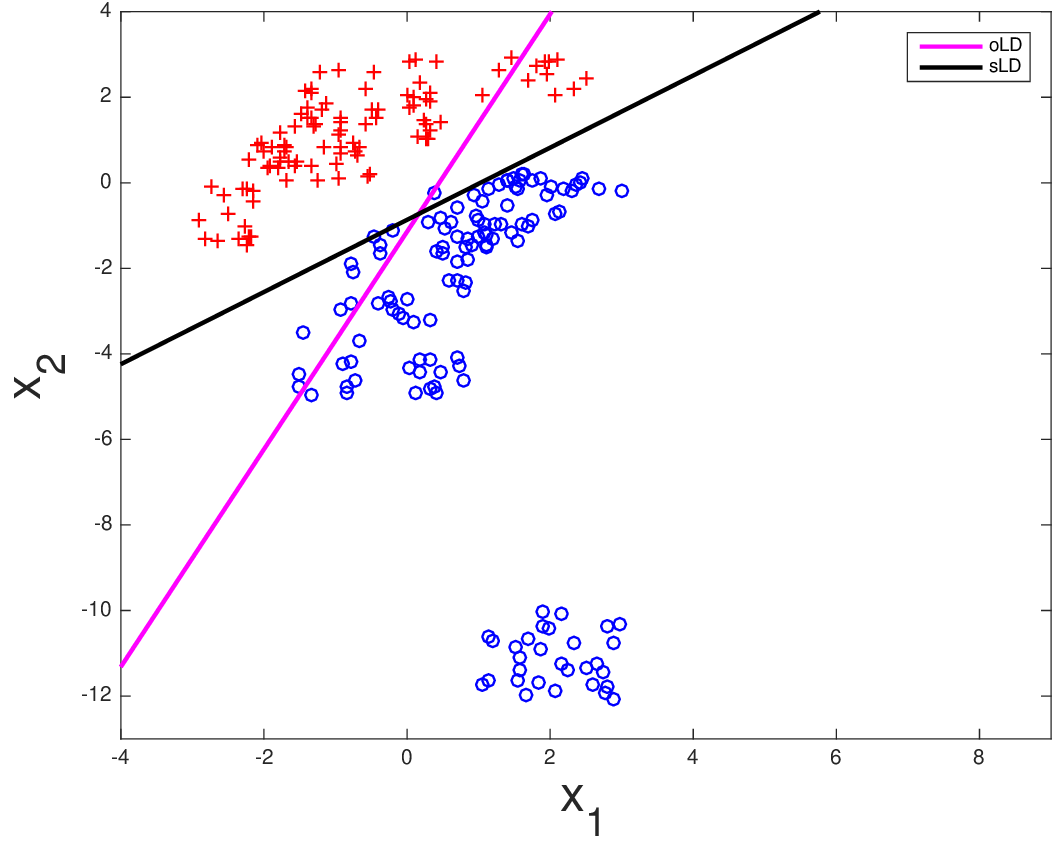} \\
\caption{Unequal density, with outlier displaced to a different position. Read text for more details.}
\label{Fig:MoreDensityandVaryPosition}
\end{figure}

Secondly, we then test our method for data density population that are not equal, namely, when there is more data in one class than the other. In this case, we considered 100 ``red crosses,'' 100 ``blue circles,'' and 30 extra ``blue circles'' outlier. In Fig.~\ref{Fig:MoreDensityAndSamePosition}, we present solution of the oLD and sLD, with the outlier still at the same position as in Fig.~\ref{Fig:Outlier},
and in Fig.~\ref{Fig:MoreDensityandVaryPosition}, we vary the position of the outlier, displacing it to a different position. While the oLD misclassifies data from both classes, it can be seen that sLD is optimal, giving a much better decision boundary.

In both Figs.~\ref{Fig:MoreDensityAndSamePosition} and \ref{Fig:MoreDensityandVaryPosition}, we see that while the oLD method is very sensitive to outliers and does poorer with change in the position of the outliers, the sLD method is very \emph{robust} against outliers and also give superior result when the position of the outliers is varied.

Lastly, we show that our method of \emph{scaled linear discriminant} does not only prove to be better than \emph{ordinary linear discriminant} when using \emph{least squares} on binary classification problems and/or synthetic data, but it is also very competitive for multiple classification and on ``real-world'' data. To this end, we tested it on the MNIST dataset for handwriting recognition. The MNIST dataset consist of $60000$ training data and $10000$ test data. Linear classifiers are generally poor on the image recognition problem, as patterns are nonlinear. Using \emph{least squares}, the accuracy of learning using \emph{ordinary linear discriminant} is $85.77\%$ on the test data, while using our method, we obtained $85.41\%$. The \emph{statistical} error in accuracy of classification using sLD relative to oLD is $\sim 0.42 \%$. Even though the oLD outperforms the sLD on the MNIST dataset by a very small margin, on some other dataset without outlier, sLD may outperform oLD as in Fig.~\ref{Fig:Normal}. But sLD may always outperform oLD in the presence of outlier, as shown in the results above. 

A note about implementation. The numerical implementation of sLD is similar to oLD, the only difference is that the augmented input vector is scaled by its norm at a cost of $\sim \mathcal{O}(ND)$ in addition to the cost of oLD method, where $D$ is the dimensionality of the input space and $N$ is the total number of samples. The formula for determining the weight vector (or matrix in multiple classification) is the same. When testing, the computed weight vector (or matrix in multiple classification) is multiplied directly with the input vectors without any further scaling applied to the input vectors.

\section{Conclusion}\label{Sec:Conclusion}
In this article, we showed a simple, effective way of improving the accuracy of linear classifiers that employ \emph{least squares} in the presence of outliers, by defining a ``scale-invariant'' linear discriminant. We presented numerical results that supported our proposition. The method also works when there are no outliers, making it more versatile than the conventional approach. 

While we have tested the method using some dataset, we do not know if there are counter examples where the method do not offer any advantage over the conventional method, or perform worse. In addition, in order to determine the full extent of the robustness and versatility of the method, it needs to be tested against other popular methods, like logistic regression, that are known to be robust against outlier.\cite{Bishop2006}

Our consideration in this article has been on data classification, whose labels take discrete values. The method presented here can be adapted to regression analysis, where ``labels'' (or the dependent variables) take continuous values. This might help provide some improvement to the \emph{outlier problem} of regression analysis when using \emph{least squares}.

\bibliography{RefLSEOutlier}

\end{document}